\documentclass{article}
\usepackage{spconf,amsmath,graphicx,hyperref}
\usepackage{xcolor}
\usepackage{amsfonts}
\usepackage{booktabs}
\usepackage{multirow}
\title{Towards Interpretable Framework for Neural Audio Codecs via Sparse Autoencoders: Exploration toward Age, Gender, and Accent Steering}
\name{\shortstack[c]{
Shih-Heng Wang, Tiantian Feng, Aditya Kommineni, Huang-Cheng Chou,\\
Bowen Yi, Xuan Shi, Shrikanth Narayanan
}}
\address{University of Southern California, USA}
\begin{document}
\ninept
\maketitle
\begin{abstract}
Neural audio codecs (NACs) are widely used in speech generation and audio-language modeling, yet how they encode speaker-trait information remains poorly understood. Prior work applied sparse autoencoders (SAEs) to investigate accent information in NACs through task-level analysis. Here, we extend this analysis to the waveform level and to age, gender, and accent, using SAE steering to probe trait-related information in sparse activations. We identify trait-associated dimensions, modify their activations, and evaluate the resulting reconstructed speech. Across five NACs, steering the selected dimensions induces target-directed shifts in speaker-trait predictions. A random-dimension baseline on Mimi produces smaller shifts, supporting the relevance of the selected dimensions. However, responses vary across codecs, traits, and steering directions, and increasing steering strength does not consistently amplify the intended shifts. Steering also generally increases word error rates and lowers predicted perceptual quality. These findings suggest that SAEs capture speaker-trait information in steerable activations, while the accompanying quality degradation highlights the need to better separate trait-related information from other information.
\end{abstract}
\section{Introduction}
\label{sec:intro}

Neural Audio Codec (NAC) representations~\cite{soundstream,encodec,dac,speechtokenizer,qwen3ttstokenizer,hificodec,wu2024ts3,moshi,espnet_codec,codecsuperb,dasb} have become a fundamental component of modern text-to-speech (TTS)~\cite{valle,valle2,vallex} and speech foundation models~\cite{moshi,audiolm,soundstorm}. Consequently, substantial effort has been devoted to benchmarking NACs~\cite{codecsuperb,dasb,espnet_codec,survey}, primarily by evaluating the effectiveness of NAC representations on downstream speech understanding and generation tasks. However, considerably less attention has been paid to understanding how these representations encode and organize information internally. This limited interpretability constrains our ability to explain and verify the behavior of NAC-based systems, particularly when they are deployed in sensitive or accountability-critical applications such as healthcare.

To address this gap, our prior work~\cite{interpretableframework} used sparse autoencoders (SAEs)~\cite{sae1,sae_claude,sae_gemma,topksae,superposition,sae_moral} to investigate the task-level interpretability of NACs for speaker accent. SAEs have been widely adopted in interpretability analysis of foundation models as they transform dense and polysemantic representations into sparse, monosemantic activations, where individual activations may correspond to human-interpretable attributes. By decomposing NAC representations using SAE, our previous study provided an initial view into how accent-related information is encoded within NAC representations. However, this analysis was limited to task-level evidence without waveform-level analysis and considered only accent, leaving its generalizability to other speaker traits unclear.

To address the first limitation, we combine SAEs with activation steering~\cite{steering1,steering2,steering3}, modifying selected SAE activations and examining the resulting changes in NAC-reconstructed speech. If modifying certain activation dimensions shifts attribute predictions for the reconstructed speech in the intended direction, it suggests that attribute-related information is decomposed into those dimensions. Prior studies have demonstrated such control in speech, audio, and music generation systems~\cite{sae_emotion,sae_codec,saemusic,sae_audio}. For example, researchers have identified steerable concepts in music generation models~\cite{saemusic} and manipulated emotion-associated SAE features to induce corresponding changes in generated speech~\cite{sae_emotion}.

Motivated by these findings, we use SAE steering as a proxy to assess whether speaker-related information in NAC representations can be decomposed into interpretable sparse activations. Note that our objective is to study the interpretability of NAC representations, rather than to develop an attribute-conversion system. To address the second limitation, we extend our analysis beyond accent to include \textbf{age} and \textbf{gender}. Together, these extensions broaden our understanding of how NAC representations encode different speaker traits.

\begin{figure*}[t]
    \centering
    \includegraphics[width=0.9\linewidth]{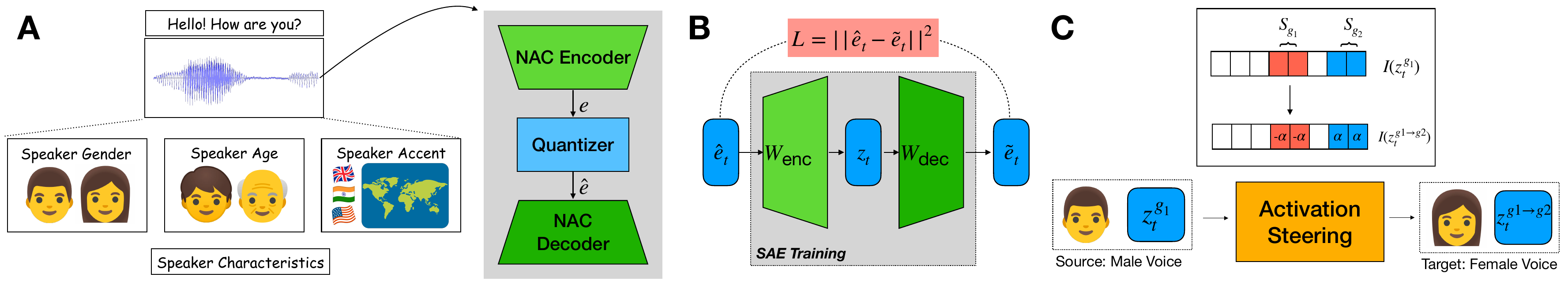}
    \vspace{-2mm}
    \caption{Overview of the SAE steering framework. \textbf{(A)} Illustration of NAC encoding speech utterance into quantized frame-level representations. \textbf{(B)} A SAE decomposes each frame into sparse activations and reconstructs it. \textbf{(C)} Trait-associated activation dimensions are identified and modified to steer reconstructed speech from a source trait $g_1$ toward a target trait $g_2$.}
    \label{fig:sae}
    \vspace{-5mm}
\end{figure*}

To achieve these goals,  we first extract NAC representations (Sec.~\ref{subsec:NAC}) and train SAEs to obtain sparse and interpretable activations (Sec.~\ref{subsec:SAE}). Given these sparse activations, we apply an activation-frequency-based algorithm to identify activations associated with specific speaker traits and perform steering on the corresponding activations (Sec.~\ref{subsec:steer}). We use Vox-Profile~\cite{voxprofile}, which aggregates 11 open-source datasets with consistent speaker-trait labels (Sec.~\ref{subsec:dataset}), and evaluate five NACs (Sec.~\ref{subsec:nac_setup}). To evaluate steering effectiveness, we use speaker-trait classification models to assess whether the predicted trait probabilities of waveforms reconstructed from the steered representations shift in the intended direction (see Sec.~\ref{subsec:trait_classification}). Our results (Sec.~\ref{sec:result}) reveal three key findings:
% To further evaluate the controllability of steering, we progressively increase the steering strength and examine whether the magnitude of the resulting speaker-trait shift increases accordingly. Last, to study the trade-off between steering effectiveness and quality degradation, we evaluate both word error rate (WER) and model subjective quality scores, which respectively measure semantic preservation and perceptual speech quality after steering.

% Importantly, our objective is not to develop a speaker-trait conversion system, but rather to use the effectiveness of SAE steering as a means of probing the interpretability of NAC representations. High-quality speaker-trait conversion is itself a challenging task, and achieving it using only SAE-based steering followed by NAC reconstruction is beyond the scope of our current setup. Instead, we aim to examine which NAC representations can be more effectively steered while better preserving speech quality under controlled interventions. A representation that supports stronger and more targeted speaker-trait manipulation with less quality degradation may indicate that the corresponding information is more readily decomposed by the SAE into sparse and potentially human-interpretable features.
\noindent \textbf{1.} SAE steering induces speaker-trait shifts, with the most consistent bidirectional control for gender and more variable responses for accent and age (Sec.~\ref{subsec:steer_result}).

\noindent \textbf{2.} Increasing steering strength can amplify these shifts, although plateaus and reversals indicate limited controllability (Sec.~\ref{subsec:steer_analysis}).

\noindent \textbf{3.} Steering generally degrades speech content and quality, suggesting that the selected activations and steering procedure do not fully isolate speaker-trait information (Sec.~\ref{subsec:steer_quality}).

% \begin{enumerate}
%     \item SAE steering induces speaker-trait shifts, with the most consistent bidirectional control for gender and more variable responses for accent and age (Sec.~\ref{subsec:steer_result}).
%     \item Increasing steering strength can amplify these shifts, although plateaus and reversals indicate limited controllability (Sec.~\ref{subsec:steer_analysis}).
%     \item Steering generally degrades speech content and quality, suggesting that the selected features and interventions do not fully isolate speaker-trait information (Sec.~\ref{subsec:steer_quality}).
% \end{enumerate}

\vspace{-3mm}
\section{SAE Steering Framework}
\vspace{-2mm}
Figure~\ref{fig:sae} summarizes our framework for identifying and steering speaker-trait-associated SAE features in NAC representations.
% To make it easier to follow, we will try to use consistent notations as our prior work~\cite{interpretableframework}.
% In Sec~\ref{subsec:NAC}, we will briefly introduce how to extract NAC features. 
% In Sec~\ref{subsec:SAE}, we will elaborate the SAE design and its interpretable features. 
% In Sec~\ref{subsec:LR}, we will elaborate how we measure the interpretability of SAE features from different NAC representation. 

\vspace{-3mm}
\subsection{Neural Audio Codec Representations}
\label{subsec:NAC}
\vspace{-2mm}
As depicted in Figure~\ref{fig:sae}(A), NACs typically consist of an encoder $\mathcal{E}$, one or more vector quantizers~\cite{vqvae}, collectively denoted by $\mathcal{V}$, and a decoder $\mathcal{D}$. 
Given a waveform $x$, the encoder produces $\mathbf{e}=\mathcal{E}(x)$, and quantization yields discrete codes and their embeddings:
$(\mathbf{c},\hat{\mathbf{e}})=\mathcal{V}(\mathbf{e})$.
Here, $\hat{\mathbf{e}} =
\left[
\hat{\mathbf{e}}_1,\ldots,
\hat{\mathbf{e}}_{T_{\text{codec}}}
\right]^\top\in\mathbb{R}^{T_{\text{codec}}\times d_{\text{codec}}}$ contains $T_{\text{codec}}$ frames.
The reconstructed waveform is $\hat{x}=\mathcal{D}(\hat{\mathbf{e}})$.

\vspace{-3mm}
\subsection{Sparse Autoencoders (SAEs)}
\label{subsec:SAE}
\vspace{-2mm}
As illustrated in Figure~\ref{fig:sae}(B), SAEs decompose dense frame-level representations $\hat{\mathbf{e}}_t$ into sparse activations $\mathbf{z}_t$. The SAE aims to learn sparse representations whose individual dimensions may correspond to human-interpretable attributes.

Following our prior work, we adopt Top-$k$ SAE~\cite{topksae}, which controls sparsity by retaining the $k$ largest activations and setting the remaining activations to zero.
The SAE consists of a linear encoder and a linear decoder with weight matrices $W_{\text{enc}}\in\mathbb{R}^{d_z\times d_{\text{codec}}}$ and $W_{\text{dec}}\in\mathbb{R}^{d_{\text{codec}}\times d_z}$, respectively, together with a shared preprocessing bias $\mathbf{b}_{\text{pre}}\in\mathbb{R}^{d_{\text{codec}}}$. No additional bias terms are used in either linear transformation.
We train the SAE model on  $\hat{\mathbf{e}}_t\in\mathbb{R}^{d_{\text{codec}}}$, treating each frame independently rather than using temporally mean-pooled utterance representations as in our prior work. Given $\hat{\mathbf{e}}_t$, its activation $\mathbf{z}_t\in\mathbb{R}^{d_z}$ and SAE reconstruction $\tilde{\mathbf{e}}_t$ are computed as
\begin{align}
\mathbf{z}_t
&=
\operatorname{TopK}
\left(
W_{\text{enc}}
\left(
\hat{\mathbf{e}}_t-\mathbf{b}_{\text{pre}}
\right),
k
\right), \label{eq:sae_e}\\
\tilde{\mathbf{e}}_t
&=
W_{\text{dec}}\mathbf{z}_t+\mathbf{b}_{\text{pre}},
\end{align}
where $d_z>d_{\text{codec}}$. The SAE is optimized using the mean squared reconstruction error
$\mathcal{L}_{\text{MSE}}
=
\lVert\hat{\mathbf{e}}_t-\tilde{\mathbf{e}}_t\rVert_2^2$.
Additional implementation details are provided in~\cite{topksae}.
Stacking the reconstructed frames yields $\tilde{\mathbf{e}}$, which is decoded into $\tilde{x}=\mathcal{D}(\tilde{\mathbf{e}})$.

\vspace{-3mm}
\begin{figure*}[t]
    \centering
    \includegraphics[width=0.88\linewidth]{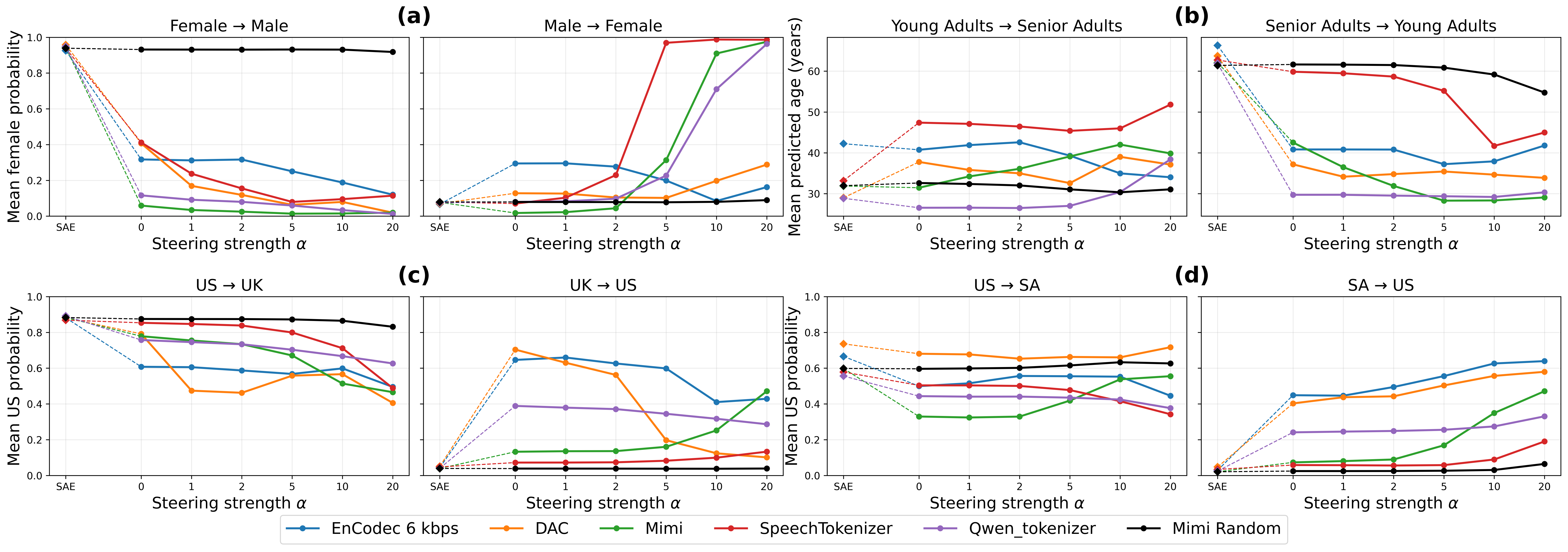}
    \vspace{-3mm}
    \caption{Steering results across five NACs for (a) gender, (b) age, (c) US--UK accent, and (d) US--SA accent. US, UK, and SA denote North American, British Isles, and Southern Asian accents, respectively. Diamonds indicate unsteered SAE reconstruction baselines; circles indicate steering results at steering strength $\alpha$. Mimi Random uses randomly selected SAE dimensions from Mimi’s representation.}
\label{fig:steering_results}
    \vspace{-6mm}
\end{figure*}

\subsection{SAE Steering}
\label{subsec:steer}
\vspace{-2mm}

As shown in Figure~\ref{fig:sae}(C), if certain dimensions of the sparse activation $\mathbf{z}_t$ capture speaker-trait information, steering those dimensions may induce corresponding changes in the SAE-reconstructed representation $\tilde{\mathbf{e}}$ and, consequently, the reconstructed waveform $\tilde{x}$. We therefore use steering as a  proxy for assessing how well the SAE decomposes such information. To perform targeted steering, we first identify trait-associated dimensions using the activation-frequency-based algorithm below:

\noindent\textbf{Activation-frequency-based algorithm.}
Following~\cite{sae_emotion}, we identify trait-associated SAE dimensions through activation frequency.
Let $\mathbf{X}_g$ denote the set of waveforms labeled with speaker trait $g$.
The activation frequency $r_i^{(g)}$ of dimension $i$ is defined as:
\begin{equation}
r_i^{(g)}
=
\frac{1}{|\mathbf{X}_g|}
\sum_{x\in\mathbf{X}_g}
\sum_{t=1}^{T_{\text{codec}}^x}
\mathbb{I}\!\left[i\in\mathcal{I}(\mathbf{z}_t^x)\right],
\end{equation}
where $\mathbf{z}_t^x$ is the SAE activation at frame $t$ of waveform $x$, $\mathcal{I}(\cdot)$ returns its active indices, $\mathbb{I}[\cdot]$ is the indicator function, and $T_{\text{codec}}^x$ is its frame count.
Unlike the utterance-level occurrence indicator in~\cite{sae_emotion}, $r_i^{(g)}$ counts repeated frame-level activations.

However, a high activation frequency $r_i^{(g)}$ at dimension $i$ alone does not establish trait specificity, as it may reflect linguistic or acoustic information shared across traits. Prior work~\cite{sae_emotion} addresses this by comparing synthesized emotional and neutral speech while holding text and speaker identity fixed. In our setting, speaker traits lack a natural neutral baseline, and real recordings do not offer the same control over content and speaker characteristics.

Therefore, we formulate a comparison between source and target trait groups, $g_1$ and $g_2$, using each group as a reference for the other to identify relative trait associations. Given their activation frequencies, we compute the normalized comparison score:
\begin{equation}
s_i^{(g_1,g_2)}
=
\frac{r_i^{(g_1)}-r_i^{(g_2)}}
{r_i^{(g_1)}+r_i^{(g_2)}+\lambda\cdot\tau}.
\end{equation}
Here, $\tau$ is the median of the nonzero values in
$\{r_j^{(g_1)}+r_j^{(g_2)}\}_{j=1}^{d_z}$,
and $\lambda$ controls stabilization to avoid small or zero denominators.
To suppress rarely activated or dead latents, we additionally set
$r_i^{(g_1)}$=$r_i^{(g_2)}$=0 for dimensions satisfying
$r_i^{(g_1)}+r_i^{(g_2)}
\leq 0.1\cdot\operatorname{mean}
\{r_j^{(g_1)}+r_j^{(g_2)}\}_{j=1}^{d_z}$.

A positive $s_i^{(g_1,g_2)}$ indicates that dimension $i$ is activated more frequently under $g_1$ than under $g_2$, and negative scores indicate the reverse. We rank dimensions in descending order of $s_i^{(g_1,g_2)}$ and $-s_i^{(g_1,g_2)}$, retaining the top $M$ from each ranking as $g_1$- and $g_2$-associated candidates for steering, respectively.

\noindent\textbf{Activation steering.}
Let $\mathcal{S}_{g_1}$ and $\mathcal{S}_{g_2}$ denote the index sets of the $g_1$- and $g_2$-associated dimensions selected above. To steer from $g_1$ toward $g_2$, we assign $-\alpha$ and $+\alpha$ to source- and target-associated dimensions, respectively, restricting changes to the original Top-$k$ active set:
\begin{equation}
z_{t,i}^{g1\rightarrow g2}
=
\begin{cases}
+\alpha,
& i\in\mathcal{S}_{g_2}\cap\mathcal{I}(\mathbf{z}_t),\\
-\alpha,
& i\in\mathcal{S}_{g_1}\cap\mathcal{I}(\mathbf{z}_t),\\
z_{t,i},
& \text{otherwise}.
\end{cases}
\end{equation}
We pass the modified activations through the SAE decoder to obtain $\tilde{\mathbf{e}}^{g1\rightarrow g2}$ and reconstruct the waveform as $\tilde{x}^{g1\rightarrow g2}=\mathcal{D}(\tilde{\mathbf{e}}^{g1\rightarrow g2})$. We perform reverse steering by exchanging $g_1$ and $g_2$, and vary $\alpha$ to assess the effect of various steering strengths.

% \textbf{}
\vspace{-3mm}
\section{Experimental setup}
\label{sec:exp_setup}
\vspace{-3mm}

\subsection{Dataset}
\label{subsec:dataset}
\noindent \textbf{Vox-Profile.}
Following our prior work~\cite{interpretableframework}, we use Vox-Profile~\cite{voxprofile} for all experiments. This benchmark aggregates 11 open-source datasets of English speech from speakers with self-reported  ages, genders, and accents, covering more than 16 English accents across regions. Its broad coverage and consistent accent annotations across sources support a comparable evaluation across speaker groups.

\noindent \textbf{Speaker traits.}
We explore age, gender and accent as the speaker traits of interest. Specifically,  we include all waveforms labeled with the following accents: \textit{North America}, \textit{English}, \textit{Welsh}, \textit{Scottish}, \textit{Northern Irish}, and \textit{Southern Asia}, together with their corresponding speaker-trait annotations.
We define source–target trait pairs $(g_1,g_2)$ as follows: {young adult} (18–30 years) vs {senior adult} (over 60 years) for age, and {female} vs {male} for gender. For accent, we examine two pairs, {US} vs {UK} and {US} vs {Southern Asia (SA)} , to assess whether our findings generalize across accent distinctions. Here, {US} denotes Vox-Profile's {North America}, {UK} denotes its {British Isles} group excluding Irish, and SA follows the corresponding Vox-Profile label. 
% The test-sample distribution for each speaker-trait pair is reported in Table~\ref{tab:test_distribution}.

\noindent \textbf{Data Splits.}
We use Vox-Profile's predefined speaker-disjoint training, validation, and test splits. The training split is used to train the SAE, and the validation split to monitor convergence. Activation analysis and steering are performed on the test split. Under our setup, selecting steering dimensions using test waveforms and their speaker-trait labels does not constitute data leakage, as both are assumed to be available to the steering procedure. The activation frequencies $r_i^{(g)}$ can be updated as new labeled data become available.

% \begin{table}[t]
%     \centering
%     \caption{Distribution of test samples across speaker-trait pairs. }
%     \include{tables/label_distribution}
%     \label{tab:test_distribution}
%    \vspace{-5mm}
% \end{table}

\vspace{-3mm}
\subsection{Neural audio codecs}
\label{subsec:nac_setup}
\vspace{-2mm}

% \noindent \textbf{NAC Models.}
Aside from EnCodec~\cite{encodec} (24\,kHz, 6\,kbps, $d_{\text{codec}}$=128), DAC~\cite{dac} (24\,kHz, $d_{\text{codec}}$=1024), SpeechTokenizer~\cite{speechtokenizer} (16\,kHz, $d_{\text{codec}}$=1024), and Mimi~\cite{moshi} (24\,kHz, $d_{\text{codec}}$=512) used in our prior work, we include Qwen3-TTS-Tokenizer-12Hz ~\cite{qwen3ttstokenizer} (24\,kHz, $d_{\text{codec}}$=512). 
% EnCodec and DAC primarily capture acoustic information, whereas SpeechTokenizer, Mimi, and Qwen\_tokenizer encode stronger phonetic structure~\cite{ssl_phonetic} through representation distillation from models such as HuBERT~\cite{hubert} and WavLM~\cite{wavlm} during pre-training. Age and gender are largely reflected in acoustic properties, while accent is more closely associated with phonetic characteristics~\cite{acc_syth}. We therefore investigate how these differences in NAC representations affect steering across speaker traits.

% \noindent \textbf{NAC model configurations.}
% For EnCodec, 24 kHz models with 1.5, 6, 12 kbps, and representation dimension $d_\text{codec}=128$ are used. For DAC, we adopt the 24 kHz version with $d_\text{codec}=1024$. For SpeechTokenizer, the sampling rate is 16 kHz with $d_\text{codec}=1024$. For Mimi, we use the 24 kHz version with $d_\text{codec}=512$.
% EnCodec (24\,kHz, 6,kbps, $d_{\text{codec}}$=128); 
% DAC (24\,kHz, $d_{\text{codec}}$=1024); 
% SpeechTokenizer (16\,kHz, $d_{\text{codec}}$=1024); 
% Mimi (24\,kHz, $d_{\text{codec}}$=512);
% Qwen\_tokenizer (24\,kHz, $d_{\text{codec}}$=512).
% For implementation, we modify the Codec-SUPERB~\cite{codecsuperb} framework to extract the NAC representation $\hat{\mathbf{e}}$.

% Note that these NACs operate at different frame rates. However, since we apply mean pooling over the time axis (see Sec~\ref{subsec:SAE}), the frame rate mismatch does not affect the resulting utterance-level representations.

\vspace{-3mm}
\subsection{Training \& Hyperparameters}
\label{subsec:sae_setup}
\vspace{-2mm}
For SAE training, official OpenAI TopK SAE implementation~\cite{topksae} was used. Following our prior work, we set the latent dimension to $d_{\text{z}}=10d_{\text{codec}}$ and the sparsity level to $k=\lceil0.1\cdot d_{\text{codec}}\rceil$, as task-level interpretability of NAC representations was observed to largely converge around these settings~\cite{interpretableframework}.

Since NACs have different $d_{\text{codec}}$, we therefore scale $d_{\text{z}}$ and $k$ with $d_{\text{codec}}$, rather than fixing them across NACs. This preserves a consistent SAE expansion ratio and relative sparsity level across NACs.
Consequently, a smaller $d_{\text{codec}}$ results in smaller $d_{\text{z}}$ and $k$. This design allows the analysis to reflect differences inherent to each representation space, where encoding reconstruction-relevant information in fewer dimensions may increase feature entanglement.

% Our design therefore controls relative SAE capacity and sparsity across NAC models while allowing differences in representational dimensionality to be reflected in their interpretability.
% The relative sparsity parameter ensures that each NAC representation retains the same percentage of active dimensions avoiding bias that could arise from using a fixed $k$, which might disproportionately favor codecs with larger or smaller $d_\text{codec}$.

% Here, the relative sparsity $s$ represents the proportion of codec dimensions allowed to remain active during reconstruction in the SAE decoder. 

We train SAE models with batch size =64 for 300 epochs. Adam optimizer with a learning rate of $1e{-5}$ and $\epsilon = 6.25e{-10}$ is used.
% The source code is available at \href{https://github.com/Stanwang1210/Speech-SAE.git}{here}.
All SAE Steering experiments use $\lambda$=100 and $M$=30.
\vspace{-4mm}
\subsection{Speaker Trait Classifiers and Baselines}
\label{subsec:trait_classification}
\vspace{-2mm}
To quantify steering effectiveness, we assess whether predicted speaker traits of steered waveforms shift toward intended targets. For age and gender, we use the 24-layer Wav2Vec~2.0~\cite{wav2vec} model from~\cite{age_gender_classification}, which jointly performs continuous age regression and gender classification and shows strong performance on Vox-Profile. For accent, we use the Vox-Profile broad-accent classifier, whose US, UK, and Other categories align with our evaluation groups; SA accents are mapped to Other. We report predicted age, female probability, and pairwise US probability for age, gender, and accent steering, respectively. To accommodate our source--target setup, we compute accent probabilities by applying softmax over the logits of the corresponding source and target classes. Table~\ref{tab:trait_baselines} reports these predictions for original and NAC-reconstructed waveforms, establishing reconstruction-induced differences before steering. Each entry is averaged over the test samples in the group specified by its column header; for example, the male-group entry is the mean female probability across male test samples. We use the same convention for the steering results.

To further assess steering effectiveness, we introduce a baseline using Mimi with random selection of $M$ SAE dimensions instead of applying our algorithm. If speaker-trait information is decomposed into specific dimensions, steering randomly selected dimensions should produce smaller trait shifts. This comparison tests whether the algorithm identifies dimensions associated with speaker traits more effectively than random selection.

\begin{table}[t]
    \centering
    \caption{Speaker-trait predictions before steering. Gender and accent scores are mean female and pairwise US probabilities (\%), respectively; age is reported in years.}
    \vspace{-2mm}
    \resizebox{\linewidth}{!}{%
\begin{tabular}{lcccc}
\toprule
\textbf{Waveform} & \textbf{Gender} & \textbf{Age} & \textbf{US--UK} & \textbf{US--SA} \\
& Female / Male & Young / Senior & US / UK & US / SA \\
\midrule
Original & 95.9 / 7.6 & 27.8 / 63.1 & 89.6 / 3.2 & 57.3 / 1.7 \\
\midrule
EnCodec 6 kbps & 94.9 / 7.8 & 36.7 / 65.0 & 89.3 / 3.8 & 67.4 / 2.4 \\
DAC & 96.0 / 7.2 & 28.7 / 63.7 & 88.2 / 5.0 & 73.7 / 4.4 \\
Mimi & 95.3 / 8.0 & 31.5 / 62.0 & 88.9 / 3.6 & 60.6 / 1.8 \\
SpeechTokenizer & 93.9 / 7.9 & 33.1 / 62.7 & 86.8 / 4.8 & 57.8 / 3.4 \\
Qwen\_tokenizer & 95.1 / 7.6 & 28.7 / 62.0 & 89.3 / 3.9 & 56.1 / 2.3 \\
\bottomrule
\end{tabular}%
}
    \label{tab:trait_baselines}
    \vspace{-8mm}
\end{table}

\begin{figure*}[t]
    \centering
    \includegraphics[width=0.88\linewidth]{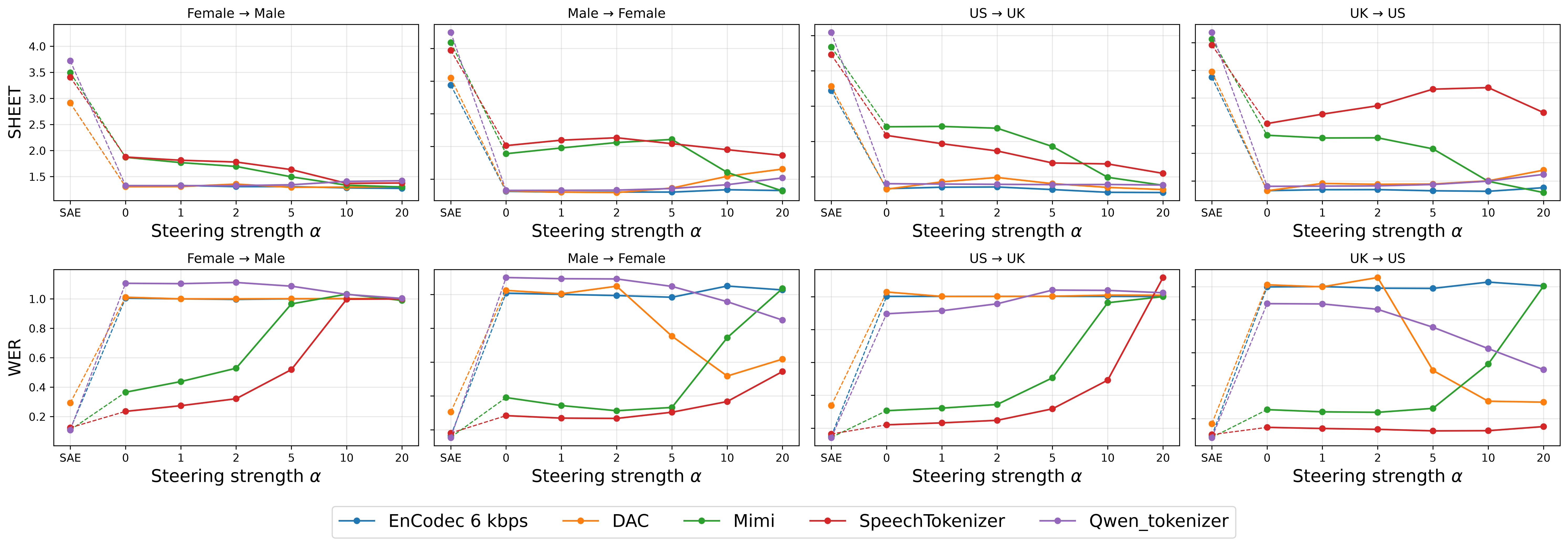}
    \vspace{-5mm}
    \caption{Speech quality after steering across NACs. The top row reports SHEET quality scores ($\uparrow$), and the bottom row reports WER ($\downarrow$). }
    \label{fig:ssqa_wer_by_alpha}
    \vspace{-5mm}
\end{figure*}

\vspace{-4mm}
\section{Result \& Analysis}
\label{sec:result}
\vspace{-3mm}

\subsection{Steering Results}
\label{subsec:steer_result}
\vspace{-2mm}
We evaluate whether SAE steering induces speaker-trait shifts in steered speech by comparing speaker-trait predictions (Sec.~\ref{subsec:trait_classification}) with those for the unsteered speech. We vary $\alpha$ from 0 to 20 to assess the direction and strength of these shifts. Note that even at $\alpha=0$, steering zeros out the selected active SAE dimensions.

\noindent \textbf{Gender.}
The Figure~\ref{fig:steering_results} (a) shows clear target-directed shifts. For female-to-male steering, female probability decreases sharply across all NACs at $\alpha$=0 and approaches close to zero as $\alpha$ increases. Mimi, SpeechTokenizer, and Qwen\_tokenizer also show strong male-to-female steering, reaching female probabilities close to 1 at $\alpha$=20. DAC and EnCodec show weaker responses in this direction. These results support the decomposition of gender information.

\noindent \textbf{Age.}
The Figure~\ref{fig:steering_results} (b) shows contrasting results for steering age. Senior-to-young steering reduces predicted age across all NACs, with sharp decreases at $\alpha$=0 for all except SpeechTokenizer. Young-to-senior steering is less consistent. SpeechTokenizer shows the largest increase, reaching 52 years at $\alpha$=20. DAC, Mimi, and Qwen\_tokenizer also exceed their SAE baselines at this strength, although their trajectories differ: Mimi rises mainly through $\alpha$=10, Qwen\_tokenizer initially shifts younger before increasing, and DAC fluctuates. EnCodec remains below its baseline throughout.

\noindent \textbf{US--UK accent.}
The Figure~\ref{fig:steering_results} (c) shows target-directed shifts in both directions. For US-to-UK steering, probability of US speaker decreases across all NACs, with the lower probabilities observed at $\alpha$=20. However, DAC and EnCodec exhibit non-monotonic shifts as $\alpha$ increases. For UK-to-US steering, EnCodec, DAC, and Qwen\_tokenizer show large increase in probability of US accent at $\alpha$=0, followed by slight decreases as $\alpha$ increases, particularly for DAC. Mimi instead shows progressively stronger shifts at higher steering strengths, while SpeechTokenizer shows slight probability increase at large steering strengths. Thus, increasing $\alpha$ does not consistently improve accent steering.

\noindent \textbf{US--SA accent.}
The Figure~\ref{fig:steering_results} (d) shows asymmetric steering, with US-to-SA steering being less successful than SA-to-US steering. For US-to-SA steering, all NACs modestly reduce US probability at $\alpha$=0, but increasing $\alpha$ does not consistently strengthen this shift. SpeechTokenizer and Qwen\_tokenizer show further reductions, whereas Mimi partially reverses its initial decrease, and EnCodec and DAC respond non-monotonically. In contrast, SA-to-US steering shows more consistent control, with US probabilities generally increasing or plateauing as $\alpha$ rises across all NACs.

\vspace{-3mm}
\subsection{Dimension Selection and Steering Strength}
\label{subsec:steer_analysis}
\vspace{-2mm}
\noindent\textbf{Validation of activation-frequency-based algorithm.}
As shown in Fig.~\ref{fig:steering_results}, steering randomly selected SAE dimensions in Mimi produces smaller target-directed shifts than steering dimensions selected by our activation-frequency-based algorithm. The contrast is clearest for gender, where random steering leaves predictions close to the SAE baseline. Random steering also yields smaller shifts for age and accent. These results suggest that the algorithm effectively identifies speaker-trait-related dimensions.

\noindent \textbf{Steering strength.}
Frequently, increasing $\alpha$ produces a stronger target-directed shift. However, the optimal value of $\alpha$ for steering effectiveness vary across NACs and speaker traits, and keep increasing $\alpha$ does not always strengthen the shift. These results provide an initial assessment of how steering strength affects SAE-based control of speaker traits in NAC representations.

\vspace{-3mm}
\subsection{Steering Quality Analysis}
\label{subsec:steer_quality}
\vspace{-2mm}
To assess whether steering modifies only speaker traits, we examine content preservation and perceptual quality after steering. Although our goal is not to develop a high-quality trait-conversion system, these metrics help assess how well the SAE separates speaker-trait information from other information. Specifically, we evaluate gender and US--UK accent steering across $\alpha$ using WER from Qwen-ASR 1.7B~\cite{qwenasr} (lower is better) and predicted perceptual quality scores from SHEET~\cite{sheet} (higher is better).

Figure~\ref{fig:ssqa_wer_by_alpha} shows that steering generally increases WER and lowers SHEET scores relative to the unsteered SAE reconstruction. Larger trait shifts often coincide with greater degradation, but the relationship varies across NACs. For example, in female-to-male steering at $\alpha$=0, Mimi and Qwen\_tokenizer produce comparable shifts in female probability, while Mimi achieves lower WER and higher SHEET scores. A similar pattern appears in US-to-UK steering. These comparisons suggest that Mimi's selected SAE features may separate trait-related information from other speech information more effectively. In general, Mimi and SpeechTokenizer preserve content and perceptual quality comparatively well in several settings, although their trait shifts are sometimes modest.

Overall, the degradation suggests that the selected SAE features do not fully isolate speaker traits from other speech information. 

\vspace{-4mm}
\section{Conclusion}
\vspace{-2mm}
In this work, we use SAE steering as a proxy to investigate how NAC representations encode speaker-trait information. Extending our prior task-level analysis of accent to waveform-level analysis of age, gender, and accent, we examine whether SAEs decompose speaker-trait information into interpretable sparse activations. Across five NACs, steering dimensions identified by our activation-frequency-based algorithm produces target-directed shifts in speaker-trait predictions. Comparisons with randomly selected dimensions in Mimi further support the relevance of the identified dimensions.

The results also reveal limits to this decomposition. Responses vary across NACs, traits, and steering directions, and increasing steering strength does not consistently amplify the intended shifts. Steering also degrades speech content and perceptual quality, suggesting that the SAE does not fully isolate speaker traits from other speech information. Future work should improve SAE optimization to better decompose information.
% /Future work should improve disentanglement and develop quality-preserving interventions, alongside perceptual evaluation to verify whether prediction shifts correspond to audible changes in the intended traits.
\clearpage
\section{Compliance with Ethical Standards}
\vspace{-3mm}
This study uses existing  datasets from Vox-Profile~\cite{voxprofile}.  Ethical approval was not required under the applicable institutional policies.
\vspace{-4mm}
\bibliographystyle{IEEEbib}
\bibliography{strings,refs}

\end{document}